**Deployment of Leader-Follower Automated Vehicle Systems for Smart Work Zone Applications with a Queuing-based Traffic Assignment Approach**

**Qing Tang**
Department of Civil and Environmental Engineering
The Pennsylvania State University, University Park, PA, 16802-1408
Email: qingtang@psu.edu

**Xianbiao Hu, Corresponding Author**
Department of Civil and Environmental Engineering
The Pennsylvania State University, University Park, PA, 16802-1408
Email: xbhu@psu.edu

## ABSTRACT

The emerging technology of the Autonomous Truck Mounted Attenuator (ATMA), a leader-follower style vehicle system, utilizes connected and automated vehicle capabilities to enhance safety during transportation infrastructure maintenance in work zones. However, the speed difference between ATMA vehicles and general vehicles creates a moving bottleneck that reduces capacity and increases queue length, resulting in additional delays. The different routes taken by ATMA cause diverse patterns of time-varying capacity drops, which may affect the user equilibrium traffic assignment and lead to different system costs. This manuscript focuses on optimizing the routing for ATMA vehicles in a network to minimize the system cost associated with the slow-moving operation.

To achieve this, a queuing-based traffic assignment approach is proposed to identify the system cost caused by the ATMA system. A queuing-based time-dependent (QBTD) travel time function, considering capacity drop, is introduced and applied in the static user equilibrium traffic assignment problem, with a result of adding dynamic characteristics. Subsequently, we formulate the queuing-based traffic assignment problem and solve it using a modified path-based algorithm. The methodology is validated using a small-size and a large-size network and compared with two benchmark models to analyze the benefit of capacity drop modeling and QBTD travel time function. Furthermore, the approach is applied to quantify the impact of different routes on the traffic system and identify an optimal route for ATMA vehicles performing maintenance work. Finally, sensitivity analysis is conducted to explore how the impact changes with variations in traffic demand and capacity reduction.

## 1. INTRODUCTION

The Autonomous Truck Mounted Attenuator (ATMA) vehicle system is an advancing technology that combines the capabilities of connected and automated vehicles (CAVs) with Autonomous Maintenance Technology (AMT) to maintain transportation infrastructure in work zones. The ATMA system employs a leader-follower approach to eliminate the need for DOT workers to be present in the follower vehicle, thus significantly reducing fatalities. The ATMA system comprises a manned leader vehicle, an unmanned follower vehicle, and a truck mounted attenuator installed on the follower vehicle [1]. The leader vehicle performs regular maintenance work, while the follower vehicle automatically duplicates the leader vehicle's actions and serves as a protective buffer in the event of a crash. Both vehicles are equipped with actuators, software, electronics, and vehicle-to-vehicle communication equipment, enabling connectivity and facilitating autonomous driving capabilities. It has gained considerable support from the Federal Highway Administration and many State DOTs, for instance, Colorado and Missouri were among the pioneering states to conduct testing and deployment of ATMA vehicles [2, 3]. Some research efforts have studied the deployment of the ATMA system. For instance, the performance of ATMA system has been evaluated and the operational design domain has been identified, aiding DOT management in determining when and where to deploy the system based on predefined traffic performance [4, 5].

However, the deployment plan of the ATMA system in a network has yet to be studied. If we assume there are multiple links that need to be maintained, and ATMA vehicles start from a single origin and will return to this same location. During maintenance operations, the slow-moving ATMA vehicles create a moving bottleneck that discounts the road capacity and leads to increased queue lengths and delays. Once the ATMA maintenance schedule is announced to the public in advance, general vehicles may choose alternative routes, resulting in different traffic assignment results based on user equilibrium (UE) principles compared to the scenario without ATMA. The presence of ATMA vehicles on different routes influences the system cost, such as the total system travel time (TSTT). The objective of this research is to identify the optimal routing of the ATMA system for transportation infrastructure maintenance, from the perspective of the entire transportation system, and to minimize the loss of efficiency.

The main challenge lies in conducting the UE traffic assignment, considering the time-varying capacity reduction, to quantify the system cost induced by ATMA operation. The user equilibrium traffic assignment problem (UETAP) has been studied since Beckmann et al. [6] introduced a mathematical program. Generally, traffic assignment methods can be classified into statistic traffic assignment (STA) and dynamic traffic assignment (DTA). The major difference between STA and DTA lies in the traffic flow models used. STA models rely on link performance functions, such as the Bureau of Public Roads (BPR) function, making it relatively easy to find UE, even for large-scale networks. The static user equilibrium traffic assignment problem (UETAP) can be mathematically formulated, ensuring the existence and uniqueness of equilibrium solutions under mild conditions. Several algorithms have been developed to solve the STA problem, for instance, the Method of Successive Averages [7], Frank-Wolfe algorithm [8], Gradient Projection Method [9] , Bush-based or origin-based algorithm [10]. STA models have been widely employed in traffic planning practice for their computational efficiency, low input requirements, and robustness, tractability, and accountability, as highlighted in [11]. However, STA models exhibit limitations in accurately representing congested condition, as they fail to account for flow metering and spillback effects [12]. Additionally, STA models assume that all vehicles on a link experience the same travel time, which does not align with real-world scenarios [13]. As such, traditional STA models are not applicable for the traffic assignment when considering the ATMA moving bottleneck.

On the other hand, significant efforts have been made in the development of DTA models to predict the evolution of traffic conditions over the last decades. DTA models offer the advantage of capturing realistic traffic flow and driver responses by tracking time-varying link flows and travel times, leading to more accurate route choice determination. Existing DTA models and algorithms can be classified into analytical-based and simulation-based approaches [14]. Among them, simulation-based methods are more widely adopted in practice due to their flexibility in network loading, and ability to simulate traffic flow propagation, capture spatial and temporal vehicular interactions, and determine link and path travel costs.

There are three methods in generating time-dependent travel times: macroscopic [15, 16], mesoscopic [17, 18], and microscopic [19, 20] models. Mesoscopic simulation approaches are commonly utilized in the network loading module, which capture changes in traffic flow at a resolution of approximately 5 to 10 seconds [21]. For instance, DYNASMART [17], DYNAMIT [18, 22], and DTALite [23]. However, DTA models generally lack neat and exact mathematical properties, resulting in lack of convergence properties that are needed for applications [21, 24]. Moreover, simulation-based models, typically designed as multiple resolution models, integrate the output of mesoscopic traffic simulators into macroscopic models. Nevertheless, this integration can introduce inconsistencies in traffic performance across different resolution levels, potentially leading to unrealistic oversaturated conditions [25]. In conclusion, DTA models do not guarantee UE or uniqueness and face challenges regarding computational efficiency, mathematical properties, and consistency in multiple resolution models.

To address these limitations, a queuing-based traffic assignment approach is proposed to provide more accuracy on congested networks when considering capacity drop caused by ATMA vehicles. To this end, based on the fluid queue model, a queuing-based time-dependent (QBTD) travel time function is introduced, in which the time-dependent queue length is determined by time-varying demand and capacity. By incorporating this QBTD travel time function into the static UETAP, we introduce dynamic characteristics and establish this traffic assignment problem. Consequently, the inclusion of a time variable in Beckmann's formulation renders the algorithms designed to solve the original UETAP inapplicable without modification. Subsequently, a path-based algorithm is modified to solve the TAP at each time step size, achieving a dynamic user equilibrium. The proposed queuing-based traffic assignment approach is then utilized to determine the optimal routing for ATMA vehicles with the lowest system cost.

The rest of this manuscript is organized as follows: Section 2 presents the preliminaries of the capacity drop induced by moving bottleneck and the QBTD travel time function. The overall module with bi-level objectives is summarized in Section 3, encompassing a higher-level objective of identifying the optimal routing of ATMA vehicles and a lower-level goal of quantifying the system cost by conducting the UE traffic assignment considering capacity drop. A path-based algorithm is then modified to solve the UETAP in each time step. In Section 4, numerical experiments are conducted to compare the performance with two benchmark models and investigate the quantitative impact of different routes on the traffic system. Sensitivity analysis is also performed to explore how the impact of ATMA vehicles changes. Section 5 concludes this paper.

## 2. PRELIMINARIES

### 2.1. ATMA Moving Bottleneck and Capacity Drop

We first introduce the capacity drop induced by the ATMA moving bottleneck. Figure 1 describes a typical highway with two lanes of the same direction, and the traffic on both lanes going to the right side. The ATMA system is represented by the two vehicles in the red box, with a leader vehicle and a follower vehicle, and a gap distance of $L_{\text{gap}}$ between these two vehicles. Other general vehicles are represented by a smaller black vehicle icon. The speed of ATMA vehicles is denoted as $v_{\text{ATMA}}$, and the cruising speed is $v_{\text{u}}$.

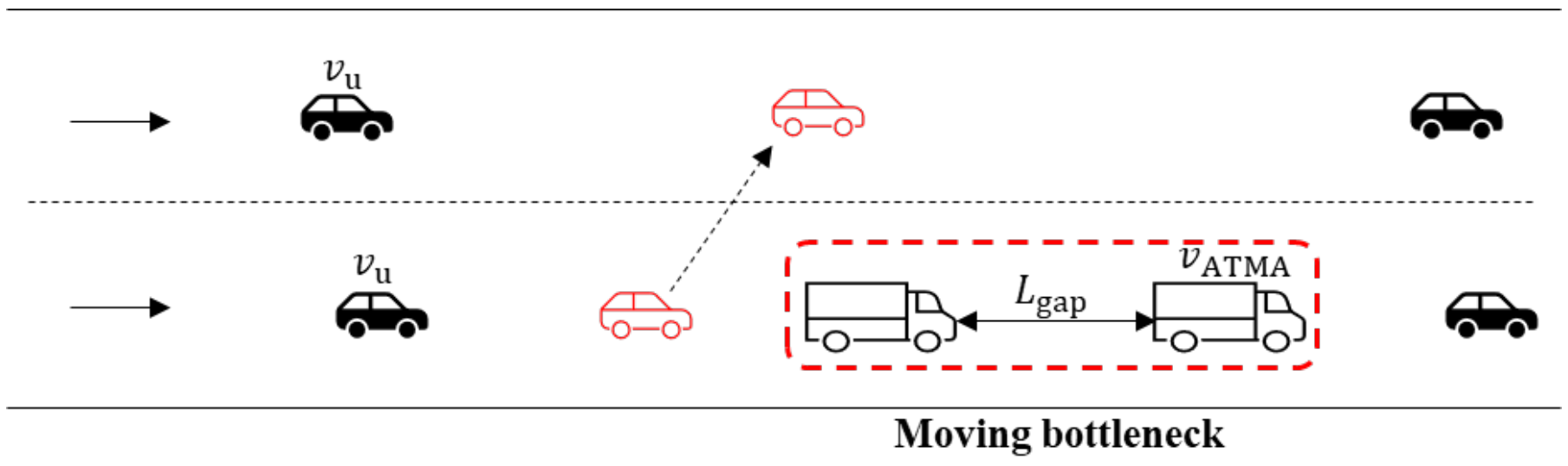

Figure 1. Schematic diagram of a four-lane highway (one direction) segment with ATMA vehicles.

The theory from Newell [26] suggests that by employing appropriate coordinate transformation, a moving coordinate system can be transformed into an analysis of flow passing a stationary bottleneck. If we position a "moving observer" on the main road, traveling at the same speed as the ATMA vehicles, the two-lane roadway segments reduce to a stationary section with only one lane from the observer's perspective. General vehicles, i.e., red vehicles in Figure 1, try to merge to one lane segment, pass this moving bottleneck, and then switch back to two lane segments. Let us call the view from the moving observer's perspective "moving coordinates", and that from a stationary observer's perspective "stationary coordinates". As such, the moving bottleneck capacity is the maximum discharge rate from a stationary observer's perspective, with the presence of ATMA vehicles. Figure 2 presents q-k fundamental diagrams (FD) of a one-lane and a two-lane roadway. The moving observer's view is marked with red (i.e., $OGI$), and the stationary observer's view is marked with black (i.e., $OG'I'$). In the moving coordinate, the triangles $OFL$ and $OGI$ represent the FD of the one-lane and two-lane segments, respectively. We use $q^*$ denote the rate at which vehicles on the main road pass the moving observer. If adding a flow $k \cdot v_{\mathrm{ATMA}}$ to $q^*$, the resulting FD will be in the stationary coordinates [26], i.e., the triangles $OF'L'$ and $OG'I'$. We have the angle of $\angle G'OI' = arctan\, v_{\mathrm{u}}$, angle of $\angle I'OI = arctan\, v_{\mathrm{ATMA}}$, and the angle of $\angle G'I'O = arctan\, w$.

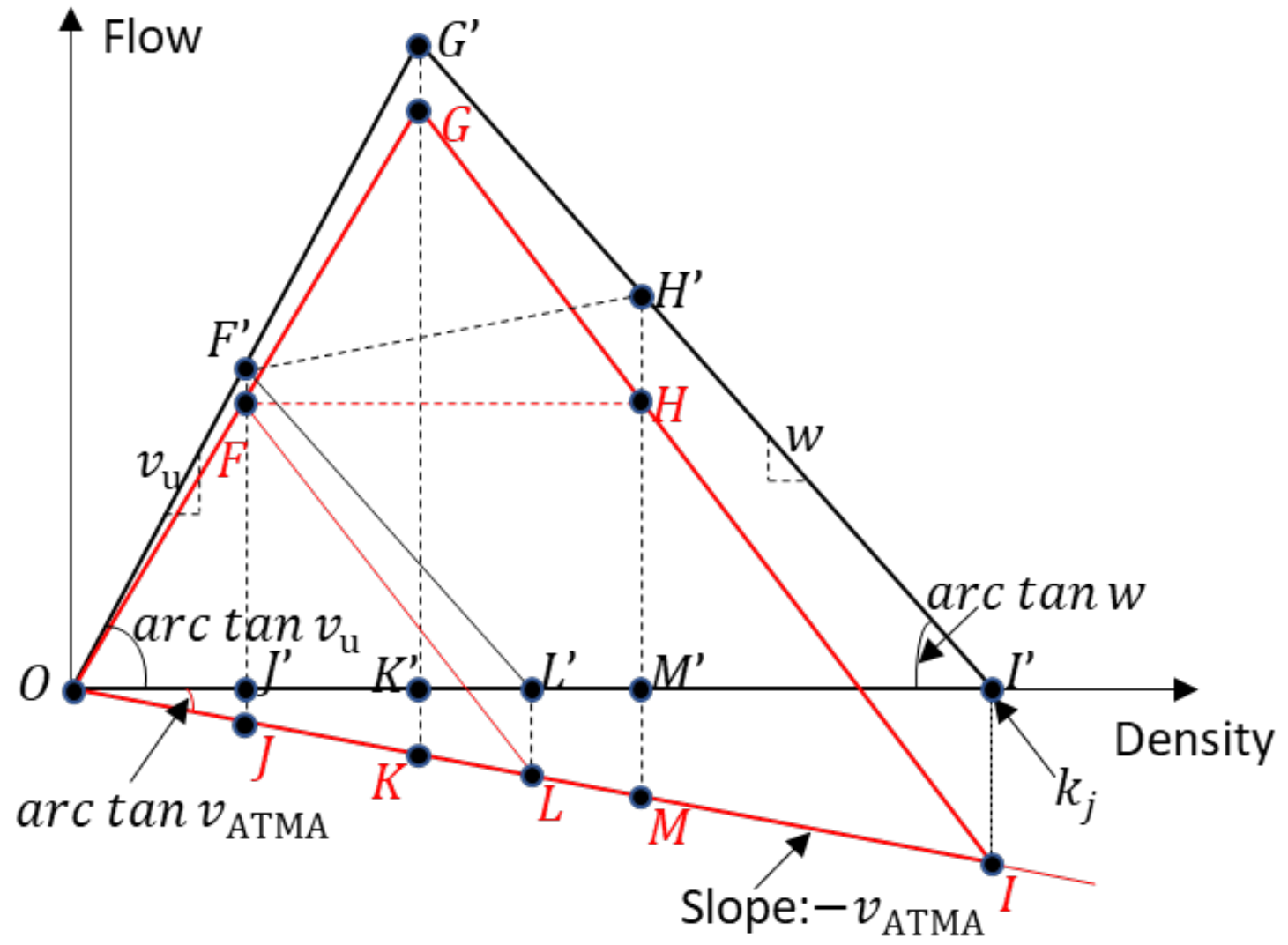

Figure 2. Flow-density relationship from both a moving observer's view and a stationary observer's view.

In the stationary coordinate system, $F'J'$ is the maximum discharge rate of the one-lane segment and $OJ'$ is its corresponding density. After a reduction by $OJ' \cdot v_{\mathrm{ATMA}}$, or $JJ'$, it becomes $FJ'$, which is the maximum discharge rate in the moving coordinate system. The traffic state of the downstream bottleneck location is represented by point $F$, while that of the bottleneck upstream is represented by point $H$, which has the same outflow rate as point $F$ but with a higher density due to the queue. When we convert the moving coordinates back to the stationary coordinate's points, points $F$ and $H$ become points $F'$ and $H'$, after adding a flow of $k \cdot v_{\mathrm{ATMA}}$. The point $H'$ determines the maximum discharge rate in the stationary coordinate system with ATMA moving bottleneck, i.e., the length of $H'M'$, which is derived by the Eq. (1). For the detailed derivation process, please refer to Tang, Hu [4].

$$\mu' = H'M' = \mu \cdot \frac{2v_{\mathrm{u}} \cdot v_{\mathrm{ATMA}} + v_{\mathrm{ATMA}} \cdot w + w \cdot v_{\mathrm{u}}}{2(v_{\mathrm{ATMA}} + w) \cdot v_{\mathrm{u}}} \tag{1}$$

where $v_{\mathrm{u}}$ is the cruising speed of general vehicles and $w$ is the backward wave speed. If we use a new variable $\theta$ and make $\theta = \frac{2v_{\mathrm{u}} \cdot v_{\mathrm{ATMA}} + v_{\mathrm{ATMA}} \cdot w + w \cdot v_{\mathrm{u}}}{2(v_{\mathrm{ATMA}} + w) \cdot v_{\mathrm{u}}}$, Eq. (1) becomes $\mu' = \mu \cdot \theta$, in which $\theta$ is, essentially, the capacity discount factor due to the ATMA moving bottleneck.

### 2.2. Queuing-based Time-dependent Travel Time Function

Next, we introduce a queuing-based time-dependent (QBTD) travel time function considering the capacity drop. During uncongested states, the aggregated vehicle speed on each road link is relatively stable and is approximated using free-flow speed [27]. While the vehicle speed will decrease due to queuing under congested states, thus, the travel time for vehicles is the summation of the free-flow travel time and the time spent in the queue or delay. The duration of queueing or delay is time-varying as the traffic demand or capacity changes. Building upon the foundation work by Newell, which employs a fluid-based approximation method to describe queue formation and dissipation process [28], we consider both uncongested and congested states. Figure 3 (a) provides an illustration of the demand-supply relationship. During peak hours from $t_0$ to $t_2$, the demand (black curve), denoted as $\lambda(t)$, exceeds the maximum discharge rate $\mu$ (red horizontal line), leading to the formation of a queue depicted in Figure 3 (b). Subsequently, the queue starts to dissipate as the arrival rate falls below $\mu$, and is fully discharged at time $\bar{t}$. The time-varying length of the queue at any given time $t$ can be calculated as:

$$Q(t) = \int_{t_0}^{t} (\lambda(\tau) - \mu)\, d\tau \tag{1}$$

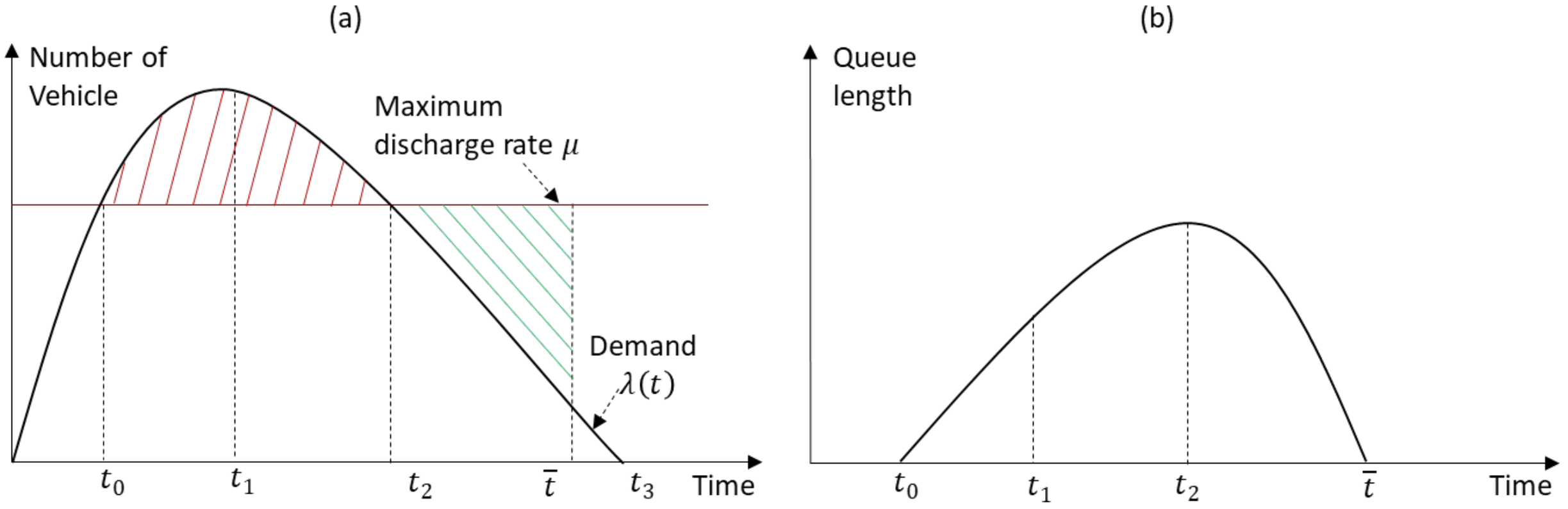

Figure 3. Queue formation and dissipation processes: (a) time-varying demand curve; (b) queuing profile.

Figure 4(a) illustrates the cumulative arrival curve $A(t)$ and departure curve $D(t)$. To describe the queue at time $t$, we shift the cumulative arrival curve by the free flow travel time to obtain a virtual arrival curve $V(t)$. Hence, the time-dependent queue length $Q(t)$ can be expressed as:

$$Q(t) = V(t) - D(t) \tag{2}$$

The horizontal distance between $V(t)$ and $D(t)$ corresponds to the time-dependent delay:

$$w(t) = \frac{Q(t)}{\mu} \tag{3}$$

As mentioned previously, the travel time consists of the free-flow travel time plus the time-dependent delay. For a general purpose, the queuing-based time-dependent (QBTD) travel time for a specific link in a network can be derived as:

$$TT(t) = t^f + w\left(t + t^f\right) \tag{4}$$

where $t$ represents the departure time or the arrival time at the upstream of a link, $t^f$ denotes the free-flow travel time for a link, $w\left(t + t^f\right)$ the delay with the departure time $t$.

**Proposition 1.** The QBTD travel time function in Eq. (4) satisfies the FIFO property.

**Proof.** The FIFO property can be proved if the inequality $t_1 + TT(t_1) \leq t_2 + TT(t_2)$ holds for any link and any time $t_1 \leq t_2$. Let $f(t) = t + TT(t)$, in which $t$ is the departure time, and $TT(t)$ is the link travel time as the departure time is $t$.

$$\frac{df(t)}{dt} = 1 + \frac{d\left(t^f + w\left(t + t^f\right)\right)}{dt} \tag{5}$$

$$= 1 + \frac{d\left(\frac{\int_{t_0}^{t+t^f} (\lambda(\tau) - \mu)\, d\tau}{\mu}\right)}{dt} = \frac{\lambda(t + t^f)}{\mu}$$

If $t_0 \leq t \leq t_2$, it is easy to observe that $\frac{df(t)}{dt} \geq 1$; otherwise, it can be found that $0 \leq \frac{df(t)}{dt} < 1$. Thus, function $f(t)$ is non-decreasing, indicating that the link travel time function $TT(t)$ satisfies the FIFO property.

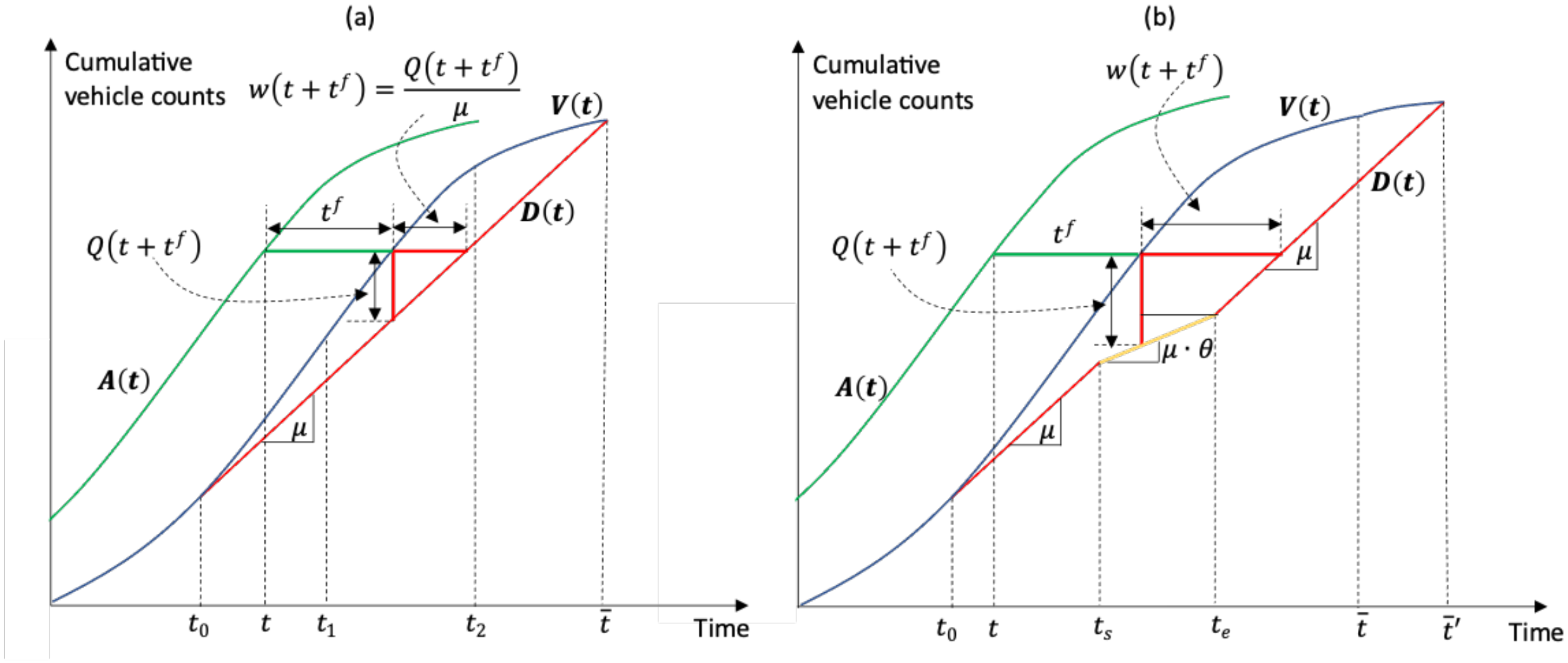

Figure 4. Cumulative curves and time-dependent travel time with (a) constant capacity and (b) discounted capacity.

As the ATMA vehicles enter the roadway segment, it creates a moving bottleneck with an effective discharge rate of $\mu'$ until the ATMA vehicles depart the roadway segment. Figure 4(b) describes the cumulative arrival and departure curves that accounts for the discounted capacity as the ATMA vehicles traverse within a specific period $[t_s, t_e]$. As such, the queue still exists at time $\bar{t}$ and is discharged at time $\bar{t}'$. The time-dependent delay time can be derived for different departure time at the upstream of a link. For example, if $t_s \leq t + t^f \leq t_e$, the delay time is derived as Eq. (6). The FIFO property is still satisfied with the corresponding QBTD travel time function with capacity drop.

$$w(t + t^f) = (t_e - t - t^f) + \frac{Q(t + t^f) - (t_e - t - t^f) \cdot \mu'}{\mu} \tag{6}$$

## 3. ATMA ROUTING OPTIMIZATION

In this section, we begin by providing an overview of the modeling framework with bi-level objectives. This framework encompasses a higher-level objective, which focuses on optimizing the ATMA route with the lowest system cost, and a lower-level objective, which involves conducting UE traffic assignment. The TAP is then formulated using QBTD travel time function and a modified path-based algorithm is developed to solve the UE problem. For clarity, a summary of the mathematical notations utilized in this paper is provided in Appendix A.

### 3.1. Overview of the modeling framework

The selection of routes for ATMA vehicles results in different time-dependent capacity reduction patterns, which significantly affect the overall performance of the traffic system. In this study, we define the system cost of a particular route, denoted as $r_{\text{ATMA}} \in R$, for ATMA vehicles as the difference between the total system travel time (TSTT) under the UE principle with and without ATMA operation, represented by Eq. (7).

$$C_{r_{\text{ATMA}}} = TSTT_{r_{\text{ATMA}}} - TSTT_0 \tag{7}$$

The optimal route, denoted as $r^*_{\text{ATMA}}$, can be obtained by minimizing the cost function: $r^*_{\text{ATMA}} = argmin\ (C_{r_{\text{ATMA}}})$. The overview of the modeling framework with bi-level objectives is depicted in Figure 5, where the higher-level objective is dedicated to determining the optimal route and the lower-level is responsible for executing the traffic assignment process. Given a potential route $r_{\text{ATMA}}$ for ATMA vehicles, the time-dependent capacity reduction can be determined and incorporated into the traffic assignment model.

As a result, the system cost, $C_{r_{\text{ATMA}}}$, associated with that specific ATMA vehicle route is obtained for each iteration loop. If the cost $C_{r_{\text{ATMA}}}$ is lower than the current minimum system cost $C_{min}$, both the minimum cost and the optimal route, $r^*_{\text{ATMA}}$, will be updated. The traffic assignment model incorporates the QBDT travel time function in the network loading module, using the given routes and route flows. The resulting time-dependent travel time and delays are utilized in the path set update and adjustment modules. The updated routes and route flows are then fed back into the network loading modules until the stop criteria are satisfied.

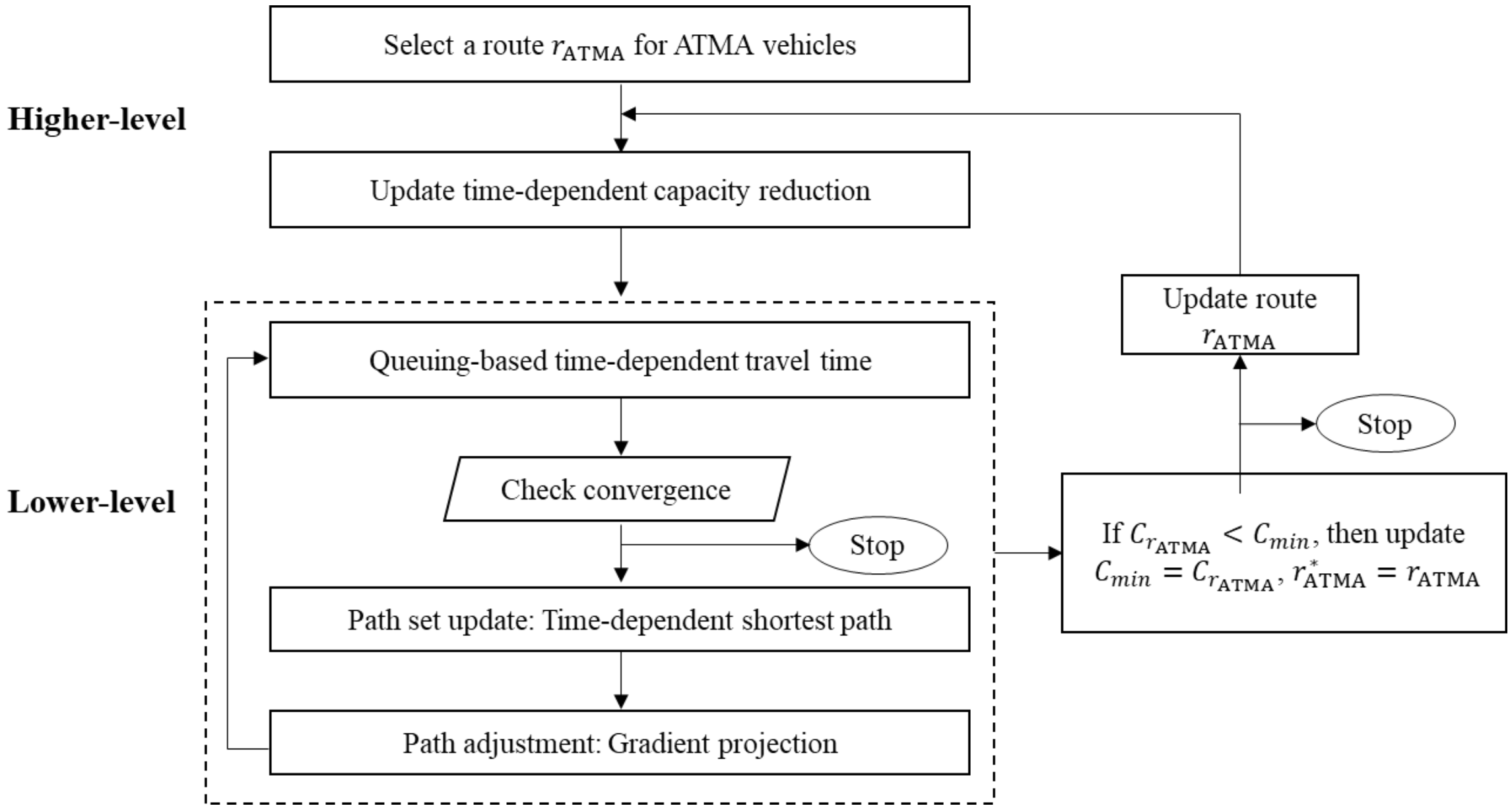

Figure 5. Overview of the modeling framework with bi-level objectives.

### 3.2. Formulation of Traffic Assignment Problem

To achieve the lower-level objective, a TAP is formulated using the QBTD travel time function at first. Before formulating this queuing-based TAP, two assumptions are made: 1) each driver wants to choose the path between their origin and destination with the shortest travel time; 2) the time-dependent link flow $x_{ij}(t)$ is continuous and differentiate.

According to Beckmann's function [6], we reformulate our queuing-based TAP utilizing the QBTD travel time function as follows:

$$\min \quad f(\boldsymbol{h}, t) = \sum_{(i,j)\in A} \int_0^{\sum_{\pi\in\Pi^{\text{rs}}} \delta_{ij}^{\pi} h^{\pi}} t_{ij}(t)\, dx \tag{8}$$

$$\text{s.t.} \quad x_{ij} = \sum_{\pi\in\Pi^{\text{rs}}} \delta_{ij}^{\pi} h^{\pi} \qquad \forall (i,j) \in A \tag{9}$$

$$\sum_{\pi\in\Pi^{rs}} h^{\pi} = d^{rs} \qquad \forall r \in O, \forall s \in D \tag{10}$$

$$t_{ij}(t) = t_{ij}^{f} + w_{ij}(t + t_{ij}^{f}) \qquad \forall (i,j) \in A \tag{11}$$

$$w_{ij}\left(t + t_{ij}^{f}\right) = \frac{Q_{ij}(t + t_{ij}^{f})}{\mu} \qquad \forall (i,j) \in A \tag{12}$$

$$Q_{ij}\left(t + t_{ij}^{f}\right) = \int_{0}^{t+t_{ij}^{f}} \left(x_{ij}(\tau) - \mu\right) d\tau \qquad \forall (i,j) \in A \tag{13}$$

$$h^{\pi} \geq 0 \qquad \forall \pi \in \Pi^{\mathrm{rs}} \tag{14}$$

The UE traffic assignment was the solution to the above optimization problem.

### 3.3. Solution Algorithm

Once we incorporate the dynamic characteristics into the traditional SUE traffic assignment, the queuing-based TAP can achieve UE at each timer interval by employing the gradient projection method, one of path-based algorithms. Different from the conventional link performance function, such as the BPR function, the travel time function described by Eq. (11) is dynamic and incorporates time $t$ as a new variable. Consequently, the traditional gradient projection method is inapplicable for addressing this queuing-based TAP. As a result, adjustments are required in the gradient projection method to effectively solve the problem.

The time-dependent Beckmann function $f(\boldsymbol{h}, t)$ is in terms of path flow $h^{\pi}$ shown in Eq. (8) and its partial derivative with respect to any path flow $h^{\pi}$ is given by Eq. (15).

$$\frac{\partial f(\boldsymbol{h}, t)}{\partial h^{\pi}} = \sum_{(i,j)\in A} \delta_{ij}^{\pi}(t) \cdot t_{ij}\left(\sum\nolimits_{\pi\in\Pi^{\mathrm{rs}}} \delta_{ij}^{\pi} h^{\pi}, t\right) = \sum_{(i,j)\in A} \delta_{ij}^{\pi} \cdot t_{ij}\left(x_{ij}, t\right) = c^{\pi}(t) \tag{15}$$

Thus, the direction of steepest descent is the negative gradient $-\nabla_{\mathbf{h}} f = -\mathrm{vect}\left(c^{\pi}(t)\right)$. We define the basic path to be a path $\hat{\pi}$ with shortest travel time, and all the other paths are named nonbasic paths. Then the constraint of capacity in Eq. (10) can be transformed into Eq. (16) by eliminating the basic path flow variable.

$$h^{\hat{\pi}} = d^{rs} - \sum\nolimits_{\pi\in\hat{\Pi}^{rs}-\{\hat{\pi}\}} h^{\pi} \tag{16}$$

Substituting Eq. (16) into the time-dependent Beckmann function (Eq. (8)), it becomes:

$$\begin{aligned} \hat{f}(\boldsymbol{h}, t) &= \sum_{(i,j)\in A} \int_{0}^{\delta_{ij}^{\hat{\pi}}(d-\sum_{\pi\in\hat{\Pi}^{rs}-\{\hat{\pi}\}} h^{\pi}) + \sum_{\pi\in\hat{\Pi}^{rs}-\{\hat{\pi}\}} \delta_{ij}^{\pi} h^{\pi}} t_{ij}\left(x_{ij}, t\right) dx \\ &= \sum_{(i,j)\in A} \int_{0}^{\delta_{ij}^{\hat{\pi}}(d-\sum_{\pi\in\hat{\Pi}-\{\hat{\pi}\}} h^{\pi}) + \sum_{\pi\in\hat{\Pi}-\{\hat{\pi}\}} \delta_{ij}^{\pi} h^{\pi}} t_{ij}(x, t)\, dx \end{aligned} \tag{17}$$

The partial derivative with respect to one of nonbasic path flow is now:

$$\begin{aligned} \frac{\partial \hat{f}(\boldsymbol{h}, t)}{\partial h^{\pi}} &= \sum_{(i,j)\in A} \left(\delta_{ij}^{\pi} - \delta_{ij}^{\hat{\pi}}\right) \cdot t_{ij}(\delta_{ij}^{\hat{\pi}}(d - \sum\nolimits_{\pi'\in\hat{\Pi}-\{\hat{\pi}\}} h^{\pi'}) + \sum\nolimits_{\pi\in\hat{\Pi}-\{\hat{\pi}\}} \delta_{ij}^{\pi'} h^{\pi'}, t) \\ &= \sum_{(i,j)\in A} \left(\delta_{ij}^{\pi} - \delta_{ij}^{\hat{\pi}}\right) \cdot t_{ij}(x_{ij}, t) = c^{\pi}(t) - c^{\hat{\pi}}(t) \end{aligned} \tag{18}$$

Let $\Delta h$ denote the amount of flow we shift away from non-basis path $\pi$ onto the basic path $\hat{\pi}$ at a time step size and let $c^{\pi}(\Delta h, t)$ and $c^{\hat{\pi}}(\Delta h, t)$ denote the travel times on path $\pi$ and $\hat{\pi}$ after shifting. Our purpose is choosing a $\Delta h$ to make the difference between $c(\Delta h, t)$ and $c^{\hat{\pi}}(\Delta h, t)$ equals to zero at each time interval. Define $g(\Delta h, t)$ as the difference travel time between a nonbasic path and basic path after shifting the flow.

$$g(\Delta h,t)=c^{\pi}(\Delta h,t)-c^{\widehat{\pi}}(\Delta h,t)=\sum_{(i,j)\in A}\left(\delta_{ij}^{\pi}-\delta_{ij}^{\widehat{\pi}}\right)\cdot t_{ij}(x_{ij}(\Delta h),t) \tag{19}$$

So

$$\frac{\partial g(\Delta h,t)}{\partial \Delta h}=\sum_{(i,j)\in A}\left(\delta_{ij}^{\pi}-\delta_{ij}^{\widehat{\pi}}\right)\cdot\frac{\partial t_{ij}(\Delta h,t)}{\partial x_{ij}}\frac{\partial x_{ij}}{\partial \Delta h}=-\sum_{(i,j)\in A_3\cup A_4}\frac{\partial t_{ij}(\Delta h,t)}{\partial x_{ij}} \tag{20}$$

where $A_3=\{(i,j)|\delta_{ij}^{\pi}=1 \text{ and } \delta_{ij}^{\widehat{\pi}}=0\}$ and $A_4=\{(i,j)|\delta_{ij}^{\pi}=0 \text{ and } \delta_{ij}^{\widehat{\pi}}=1\}$.

Based on the assumption that the link flow is continuous and differentiate, Newton's method is adopted to estimate the shifted flow $\Delta h$ at each time interval.

$$\Delta h=0-\frac{g(0,t)}{g'(0,t)}=\frac{c^{\pi}(\Delta h,t)-c^{\widehat{\pi}}(\Delta h,t)}{\sum_{(i,j)\in A_3\cup A_4}\frac{\partial t_{ij}(\Delta h,t)}{\partial x_{ij}}} \tag{21}$$

The amount of flow to shift is then:

$$\Delta h=\min\{h_{\pi},\frac{c^{\pi}(\Delta h,t)-c^{\widehat{\pi}}(\Delta h,t)}{\sum_{(i,j)\in A_3\cup A_4}\frac{\partial t_{ij}(\Delta h,t)}{\partial x_{ij}}}\} \tag{22}$$

Next, we need to derive $\frac{dt_{ij}}{dx_{ij}}$ to solve our proposed semi-dynamic UE traffic assignment problem. If the problem is formulated in continuous time, we can derive the first derivation of travel time function to the link flow as Eq. (23). For the derivation process, please refer to Appendix B.

$$\frac{\partial t_{ij}\left(x_{ij},t\right)}{\partial x_{ij}}=\frac{x_{ij}(t)}{\mu_{ij}(t)}/\frac{dx_{ij}(t)}{dt} \tag{23}$$

Since the link flow $x_{ij}$ is unknown before we perform traffic assignment, it is impossible to derive the first derivation unless an exact function is given. Moreover, solving continuous-time models in large-scale networks could be challenge and hence numerical solution algorithm are usually developed based on discretized time interval (e.g., Ziliaskopoulos [29], Qian, Shen [30], and Long, Chen [31]). To simplify the proximation process of the first derivation of $\frac{dt_{ij}}{dx_{ij}}$, we take discretized time interval in this manuscript. The analysis period is denoted as $T$, with a time interval set as $\Delta t$. Therefore, the total number of time intervals is $M=T/\Delta t$ with each interval $m=\{1,2,\dots,M\}$. We then assume that the flow on any link $(i,j)$ is a constant for each $k^{th}$ time interval $\Delta t$. In this case, the queue length at $m^{th}$ time interval becomes:

$$Q_{ij}(m\cdot\Delta t)=\int_{(m-1)\cdot\Delta t}^{m\cdot\Delta t}\left(x_{ij}^{m}-\mu_{ij}\right)d\tau \tag{24}$$

The time-dependent delay time can be calculated as

$$w_{ij}(m\cdot\Delta t)=\frac{Q_{ij}(m\cdot\Delta t)}{\mu_{ij}}=\frac{\int_{(m-1)\cdot\Delta t}^{m\cdot\Delta t}\left(x_{ij}^{m}-\mu_{ij}\right)d\tau}{\mu_{ij}} \tag{25}$$

Then the QBTD travel time function on link $(i,j)$ is

$$\begin{aligned}\left(x_{ij},m\cdot\Delta t\right)&=t_{ij}^{f}+w_{ij}\left(m\cdot\Delta t+t_{ij}^{f}\right)=t_{ij}^{f}+\frac{\int_{(m-1)\cdot\Delta t+t_{ij}^{f}}^{m\cdot\Delta t+t_{ij}^{f}}\left(x_{ij}^{m+t_{ij}^{f}/\Delta t}-\mu_{ij}\right)d\tau}{\mu_{ij}}\\&=t_{ij}^{f}+\frac{\left(x_{ij}^{m+t_{ij}^{f}/\Delta t}-\mu_{ij}\right)\cdot\Delta t}{\mu_{ij}}\end{aligned} \tag{26}$$

We can derive the $\frac{dt_{ij}}{dx_{ij}}$ as Eq. (27), and for the derivation process, please refer to Appendix B.

$$\frac{dt_{ij}\left(x_{ij}, m\cdot\Delta t\right)}{dx_{ij}} = \begin{cases} \frac{\Delta t}{\mu_{ij}(m\cdot\Delta t)}, & \text{if } x_{ij} > \mu_{ij} \\ 0, & \text{if } x_{ij} < \mu_{ij} \end{cases} \tag{27}$$

To outline the procedure of the queuing-based traffic assignment, we first introduce some notations. An origin-destination (OD) pair is represented by $n \in \{1,2,\dots,N\}$ with a path set denoted as $\Pi_n$ and individual paths within the set as $\pi_n \in \Pi_n$. The travel time for each path is denoted as $c_{\pi_m}$. The solution algorithm for this traffic assignment is detailed in the following.

**Description of the solution algorithm**

1. Initialization: $m = 1$, $\widehat{\Pi}_n = \emptyset$, $Q_{ij}(0) = 0$, $\mu_{ij}(0) = 0$, $d_n(0) = d_n$, $x_{ij}(0) = 0$, $tt_{ij}(0) = t_{ij}^f$, $c_{\pi_n} = \sum_{(i,j)\in\pi_n} tt_{ij}$

2. $t = m\cdot\Delta t$ , update $\mu_{ij}(m\cdot\Delta t)$ and $d_m(m\cdot\Delta t)$

2.1. $n = 1$

2.1.1. Iteration $k = 1$

2.1.2. Find the shortest path $\pi_n^*$, if not in $\widehat{\Pi}_n$ adding it.

2.1.3. If there is only one path in $\widehat{\Pi}_n$, set the path flow $h_{\pi_n^*} = d_n(m\cdot\Delta t)$; otherwise, for each nonbasic path $\pi_n$ calculate shift flow $\Delta h_{\pi_n} = \frac{c_{\pi_n} - c_{\pi_n^*}}{\sum \frac{dtt_{ij}}{dx_{ij}}}$. Set $h_{\pi_n^*}^k = h_{\pi_n^*}^{k-1} + \sum \Delta h_{\pi_n}$, and $h_{\pi_n}^k = h_{\pi_n}^{k-1} - \Delta h_{\pi_n}$. Remove path $\pi_n$ if $h_{\pi_n}^k \le \varepsilon$

2.1.4. Update link flow $x_{ij}^k(m\cdot\Delta t)$

2.1.5. Calculate link travel time $tt_{ij}^k(m\cdot\Delta t) = t_{ij}^f + \frac{\max\left(0, Q_{ij}((m-1)\cdot\Delta t) + \Delta t\cdot\left(x_{ij}^k(m\cdot\Delta t) - \mu_{ij}((m-1)\cdot\Delta t)\right)\right)}{\mu_{ij}((m-1)\cdot\Delta t)}$

2.1.6. Update path cost $c_{\pi_n}^k = \sum_{(i,j)\in\pi_n} tt_{ij}^k(m\cdot\Delta t)$

2.1.7. Relative gap $\gamma \le \epsilon$ or $k \ge iter_{max}$?

- If yes, check $n \ge N$?
  - If yes, check $t \ge T$?
    - If yes, output $TSTT$ and stop.
    - If no, update $Q_{ij}(m\cdot\Delta t), m = m + 1$, return to Step 2.
  - If no, $n = n + 1$, return to Step 2.1.1
- If no, $k = k + 1$, return to Step 2.1.2

## 4. NUMERICAL EXPERIMENTS

To validate and assess the queuing-based traffic assignment approach, we conducted numerical experiments on two different network sizes: a small-size simple network and the Sioux Falls network. Initially, we verified the convergence pattern of our proposed algorithm and assessed the benefits of incorporating capacity drop modeling and the QBTD travel time function in UE traffic assignment. Additionally, we utilized the queuing-based approach to analyze the impact of ATMA vehicles on the overall traffic system and identify the optimal route with the lowest system cost. Sensitivity analysis was also performed to examine how the results vary with changes in demand and the travel speed of ATMA vehicles. The experiments were conducted using Path4GMNS [32], an open-source platform for UE traffic assignment, which was adapted and modified to accommodate our queuing-based TAP.

### 4.1. Methodology Validation and Benefit Analysis

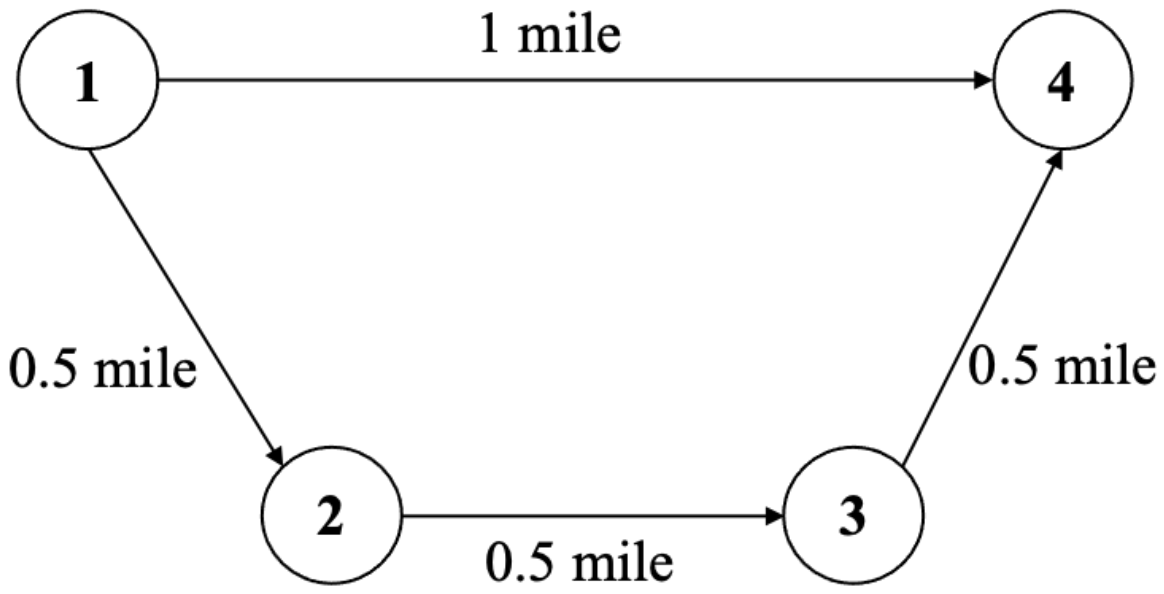

Figure 6 illustrates the topology of the small-size network. The traffic demand from the origin node 1 to the destination node 4 is 6,000 veh/hr. Two paths are available: P1 (1,4) and P2 (1,2,3,4). For each link, the free flow travel speed is 40mph, and the backward wave speed is 12mph. Additionally, the capacity of each link is 3,000 veh/hr. The first path has a free flow travel time of 90 s, while the second path has a free flow travel time of 135 s.

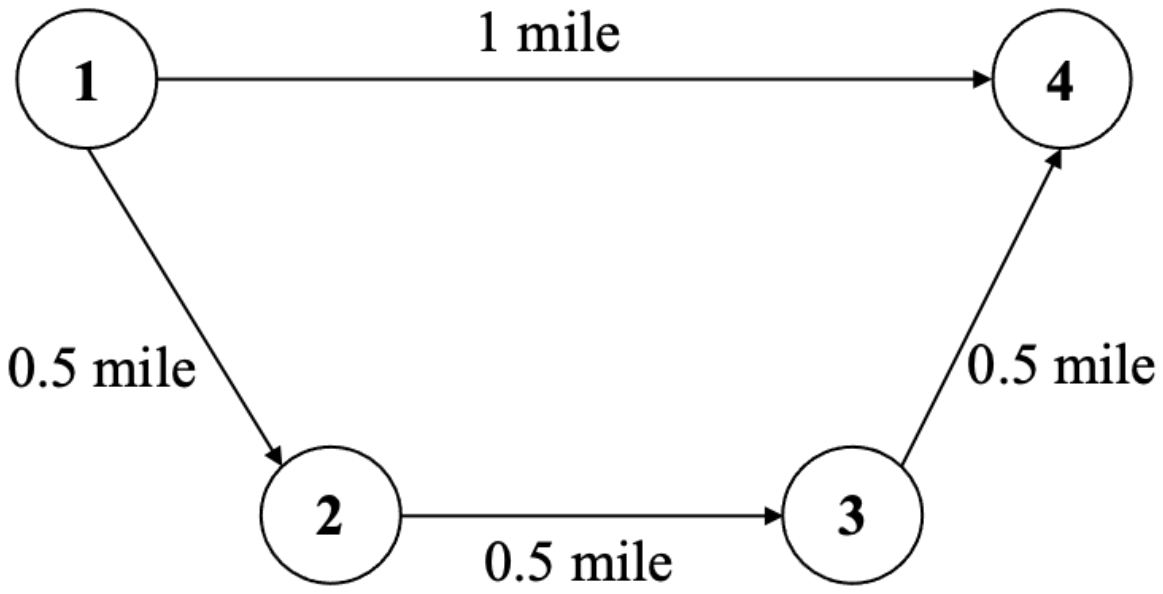

Figure 6. The simple network topology.

Initially, we verified the convergence pattern of the adjusted gradient projection algorithm, which incorporates the QBTD travel time function. In the algorithm, we examined various time intervals ranging from 5s to 1min to analyze the convergence pattern. Figure 7 illustrates the convergence pattern as the time interval increases. Notably, we observed that it requires 7 iterations to converge to 0.01% when the time interval is set to 5s, while it requires 28 iterations if the time interval is1 min. As the time interval increases, the initial relative gap also increases, consequently necessitating more iterations to achieve convergence.

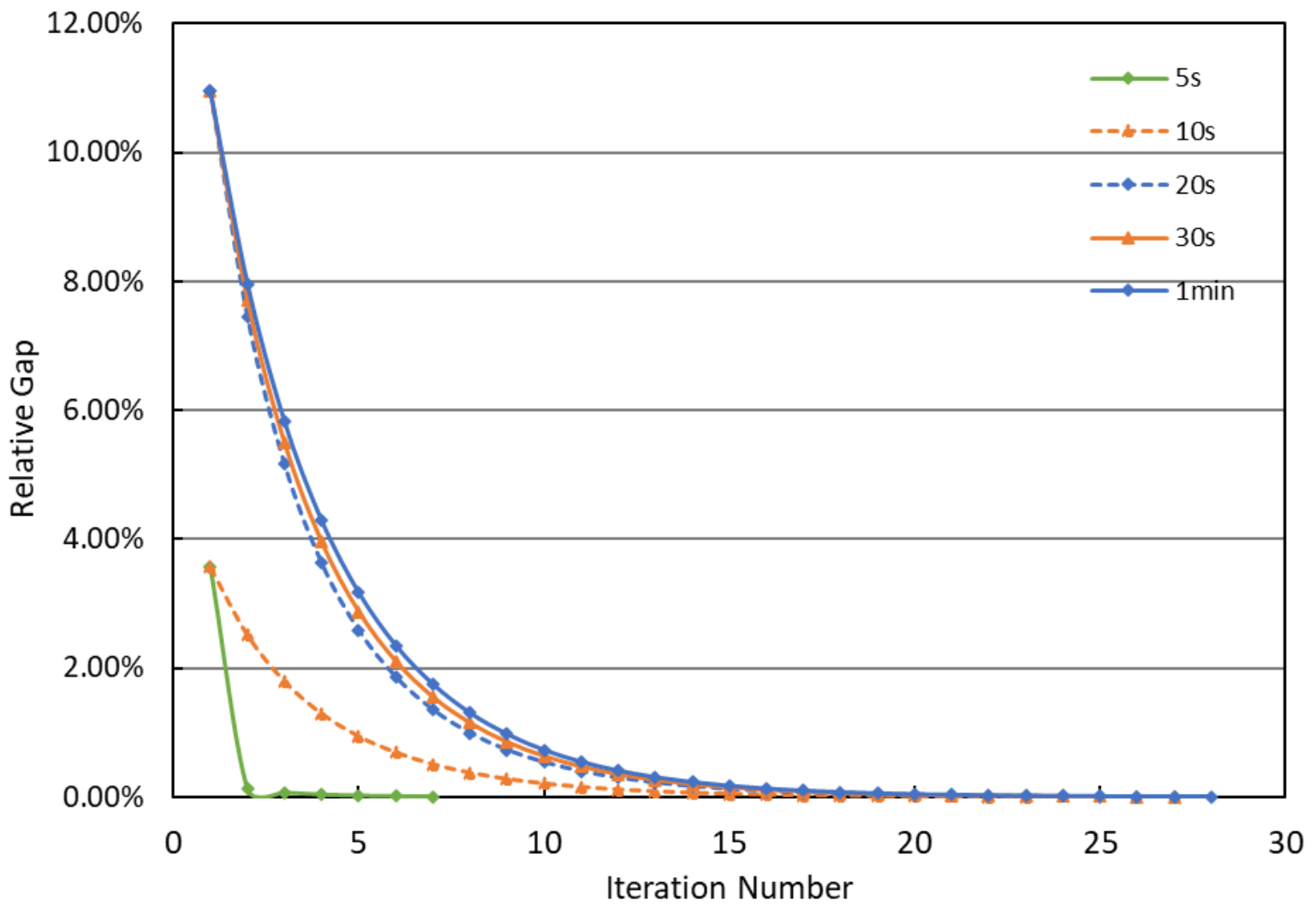

Figure 7. Convergence pattern.

Next, we conducted experiments to evaluate the performance of the proposed queuing-based traffic assignment approach. For each time interval, we employed the adjusted gradient projection method to perform UE traffic assignment. In this algorithm, we incorporated the QBTD travel time function, considering capacity drop. To assess the benefits of our proposed model, we established two benchmark models. In the first benchmark model, capacity drop was ignored in the QBTD travel time function during UE traffic assignment. However, realistic capacity was considered to determine actual travel time and relative gap. By comparing the proposed model with this benchmark model, we could analyze the benefits of capacity drop modeling. In the second benchmark model, the BPR travel time function considering capacity drop was employed for UE traffic assignment in each time interval. While the QBTD travel time function, accounting for capacity drop, was used to calculate the actual travel time and relative gap. By comparing the proposed model with the second benchmark model, we could observe the advantages of using the QBTD travel time function. In the experiment, the total time was set to 600s with a time interval of 30s. We assumed that the link (1,4) required maintenance, which was carried out by slow-moving ATMA vehicles with a speed of 3.5 m/s. The total maintenance time was 460s, during which the effective discharge rate on link (1,4) decreased. Once the maintenance was completed, the link's capacity returned to normal. For each time interval, a UE traffic assignment was performed, and the queuing length and travel time of the link were updated accordingly. Figure 8 presents a comparison of the relative gaps between the proposed model and the two benchmark models. The blue line represents the relative gap of the proposed model, which remains close to 0.0%. It can be observed that the corrected relative gap of benchmark model 1 (depicted by the orange line) gradually increases to 60% during the capacity reduction period and then drops to around 40% after the capacity returned to normal. This implies that if we disregard the capacity drop in the UE traffic assignment, the resulting mismatched path flow will deviate from UE in this simple network. Furthermore, when utilizing the static BPR travel time function in the UE traffic assignment, the corrected relative gap (illustrated by the green line) exhibits a continuing increase up to 82%. It indicates that if the congested condition is not captured, the resulting path flow significantly deviates from the equilibrium.

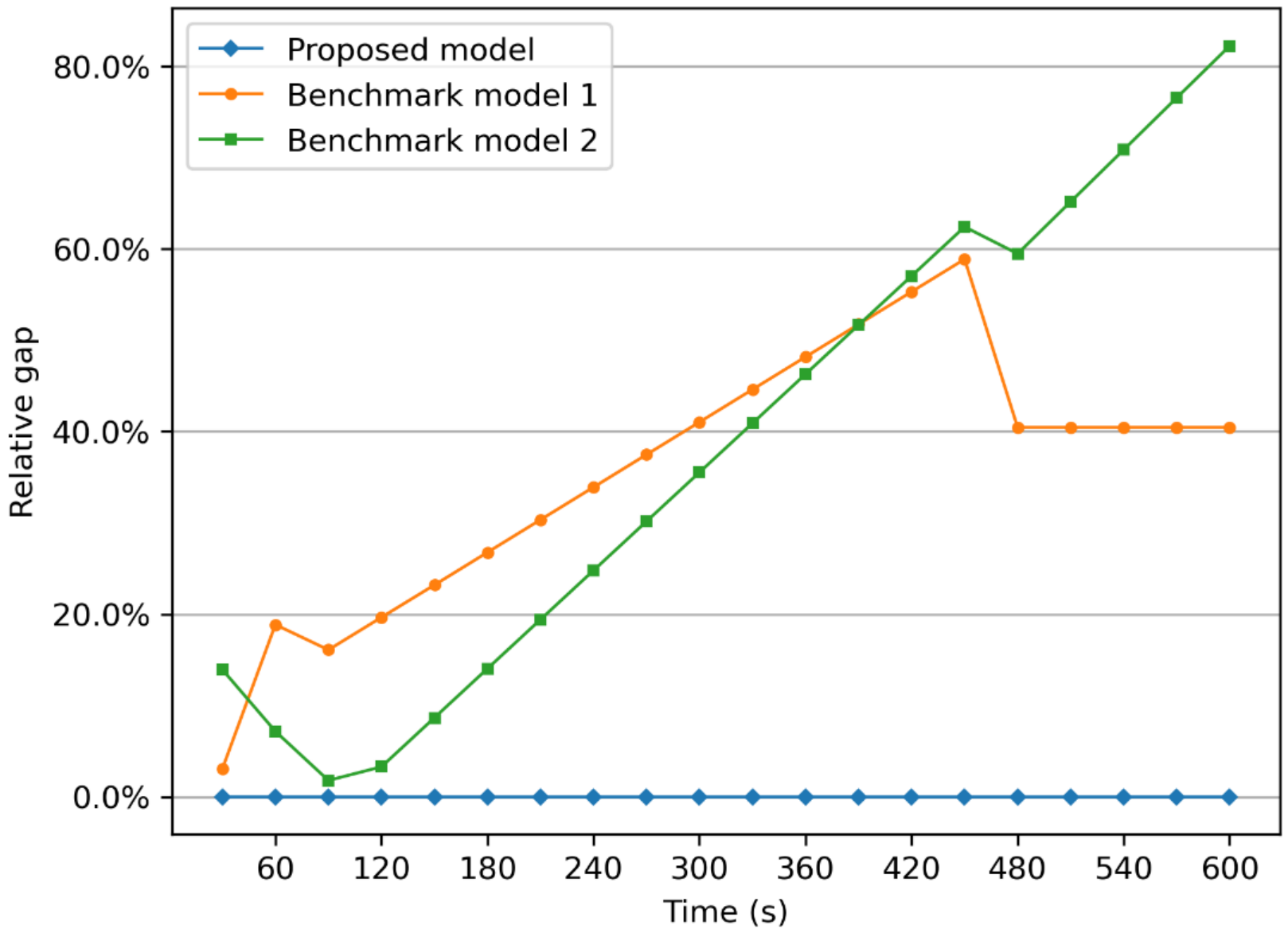

Figure 8. Relative gap comparison of proposed model and two benchmark models.

### 4.2. Routing Plan Benefit Analysis

This section focuses on examining the benefits of the proposed model in the routing plan for ATMA vehicles. In a network, multiple links may require simultaneous maintenance, which poses a challenge in finding the optimal route for ATMA vehicles. Typically, the shortest path is considered the best choice. However, the slow-moving ATMA vehicles become a moving bottleneck, leading to capacity drops. Consequently, the link queue length and travel time increase during the maintenance period. For different routes, it is necessary to determine the time-varying capacity for specific links. Our proposed model incorporates capacity reduction in UE traffic assignment and quantifies the impact of various ATMA routes on the traffic system. The obtained results will be compared with those of the two benchmark models to evaluate the effectiveness of our approach.

#### *4.2.1. Convergence Pattern*

To illustrate the impact of different routes, we will consider the Sioux Falls network [33], as depicted in Figure 9. This network consists of 24 nodes and 75 links. The free flow travel speed is 60mph, while the backward wave speed is 20mph. In our analysis, we will focus on the maintenance of four specific links: (6,8), (16,17), (15,22), and (11,14), which are represented by the red links in Figure 9. The travel speed of ATMA vehicles during maintenance is set to 10mph. There are multiple routes available for ATMA vehicles with different origins and destinations. In this study, we select node 6 as the origin and node 14 as the destination. Specifically, we consider the impact of the shortest 10 paths on TSTT. For instance, the free flow travel time (FFTT) for general traffic of the shortest path is 38 min, whereas for ATMA vehicles, it increases to 228min. The FFTT for the other nine paths ranges from 39min to 48min for general traffic, and from 234min to 288min for ATMA vehicles.

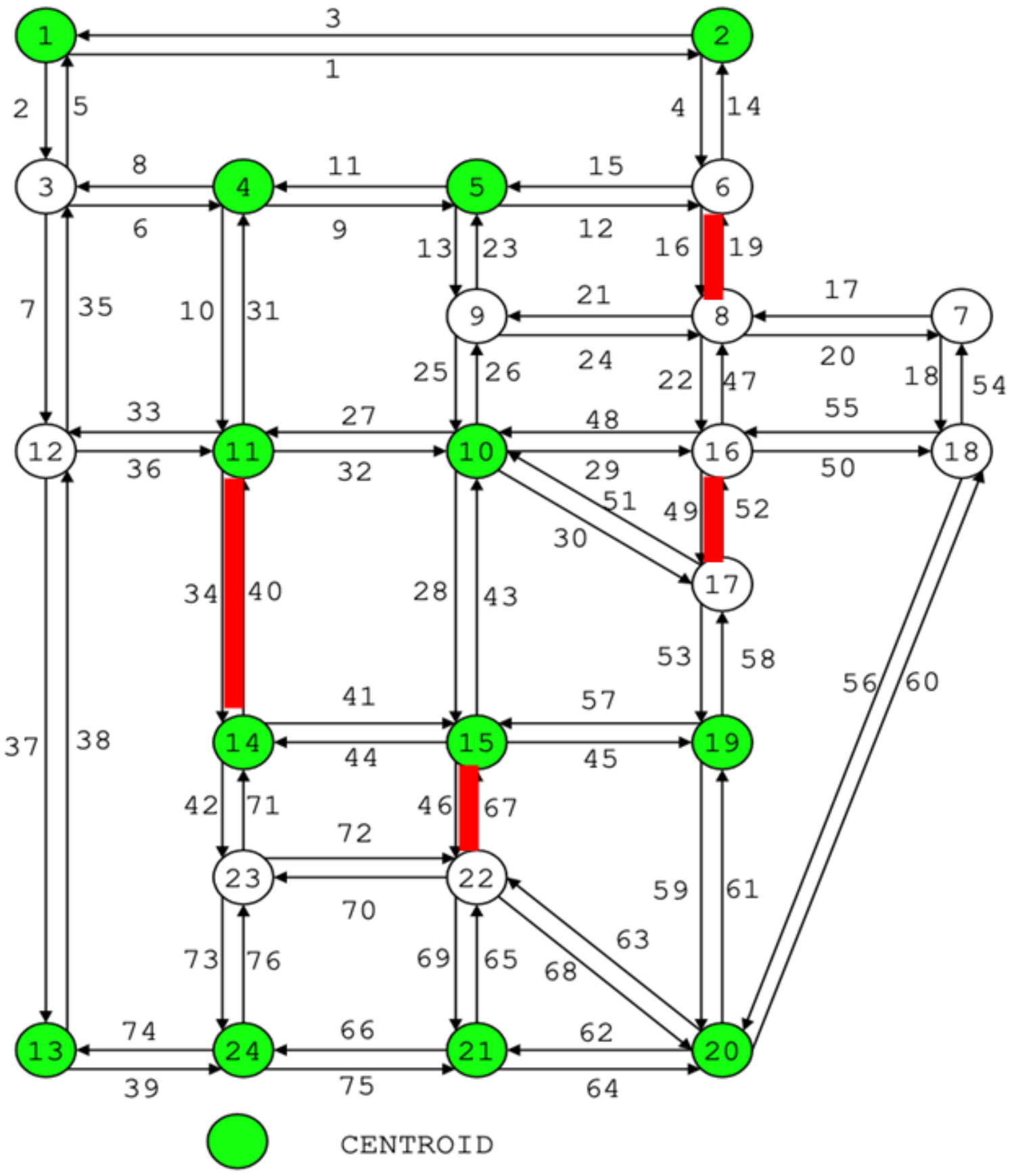

Figure 9. Sioux Falls Test Network.

The experiment was conducted over a total duration of 5 hours, with a time interval of 5 seconds, resulting in 3,600 intervals in total. Convergence criteria were established, requiring either a maximum of 20 iterations or a relative gap below 0.1% for each time interval. During our experimental analysis, we thoroughly investigated the ten different paths characterized by varying capacity drop patterns. The results of our study exhibited an impressive level of convergence, with relative gaps below 0.1% observed in 98.9% to 99.1% of all time intervals across the examined paths. For instance, Figure 10 presents a cumulative histogram of relative gap. These results demonstrate the effectiveness of the proposed queueing-based traffic assignment and the modified path-based algorithm in achieving convergence.

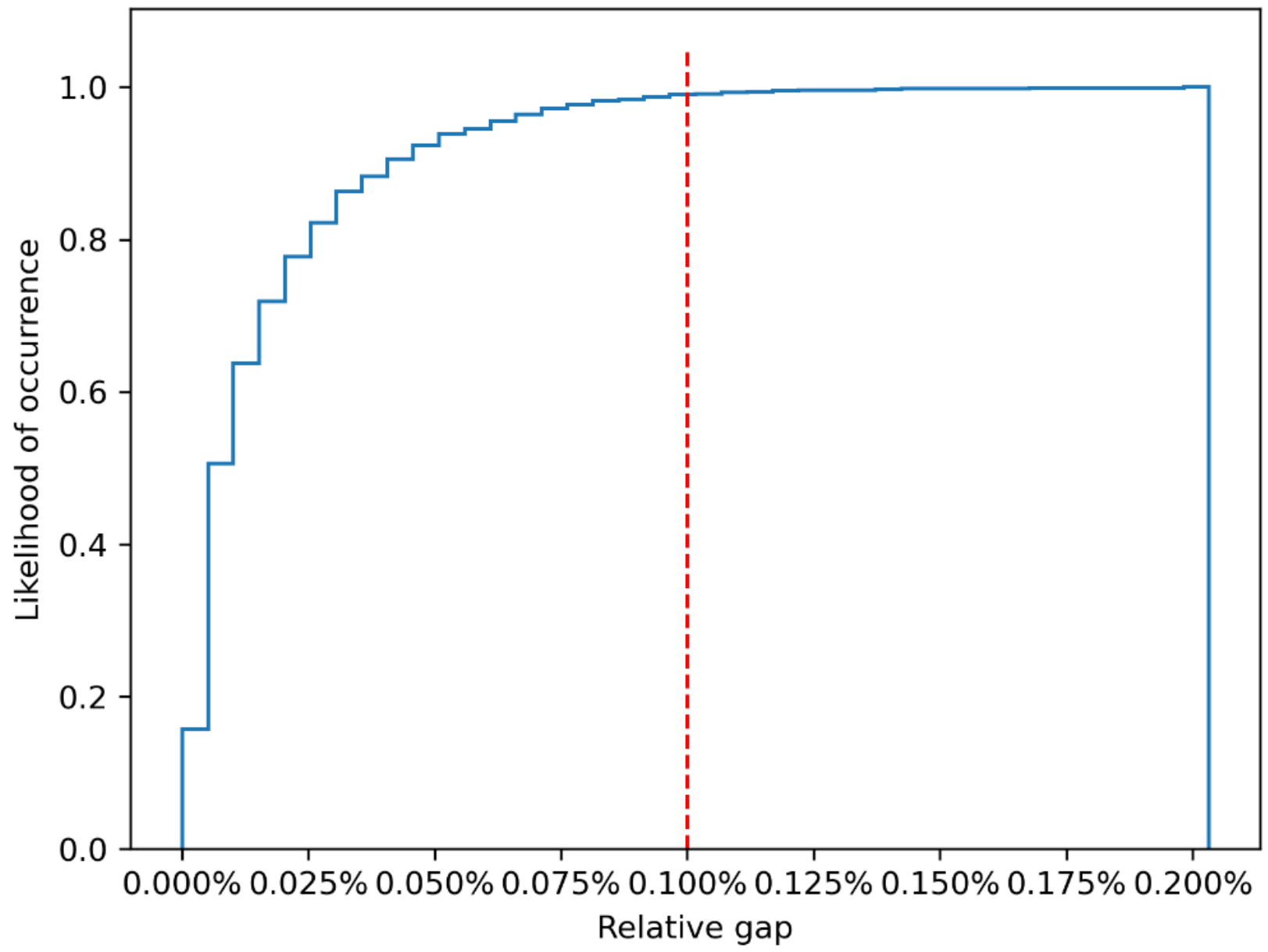

Figure 10. Cumulative histogram of relative gap.

*4.2.2. Benefit Analysis*

To evaluate the benefits of the capacity drop modeling and the QBTD travel time function, we conducted experiments for each path using the proposed model and two benchmark models. Figure 11 presents cumulative histograms comparing the corrected relative gaps of the benchmark models with those of the proposed model. The results reveal significant deviations from the UE in the benchmark model 1, with corrected relative gaps exceeding 0.1% observed in 90.4% to 99.1% of all time intervals across the examined paths when the capacity reduction is disregarded. Additionally, when the UE traffic assignment utilizes the BPR function instead of the QBTD travel time function, the results demonstrate substantial deviations from the UE, with corrected relative gaps exceeding 2% observed in 98.9% of all time intervals across the examined paths. This highlights the inadequacy of the static BPR travel time function in capturing time-varying information and its consequent significant deviation from the user equilibrium.

Table 1 provides a summary of the average corrected relative gaps for the 10 paths between the proposed model and the two benchmark models. The average relative gap for the proposed model is approximately 0.019%. However, if the capacity reduction is disregarded in the UE traffic assignment process, the average relative gap increases to a range of 0.286% to 0.582%. Additionally, when using the static BPR travel time function, the average relative gap sharply increases to a range of 11.71% to 11.86%. These results highlight the substantial impact of considering capacity drop and utilizing the QBTD travel time function in achieving a closer approximation to the user equilibrium.

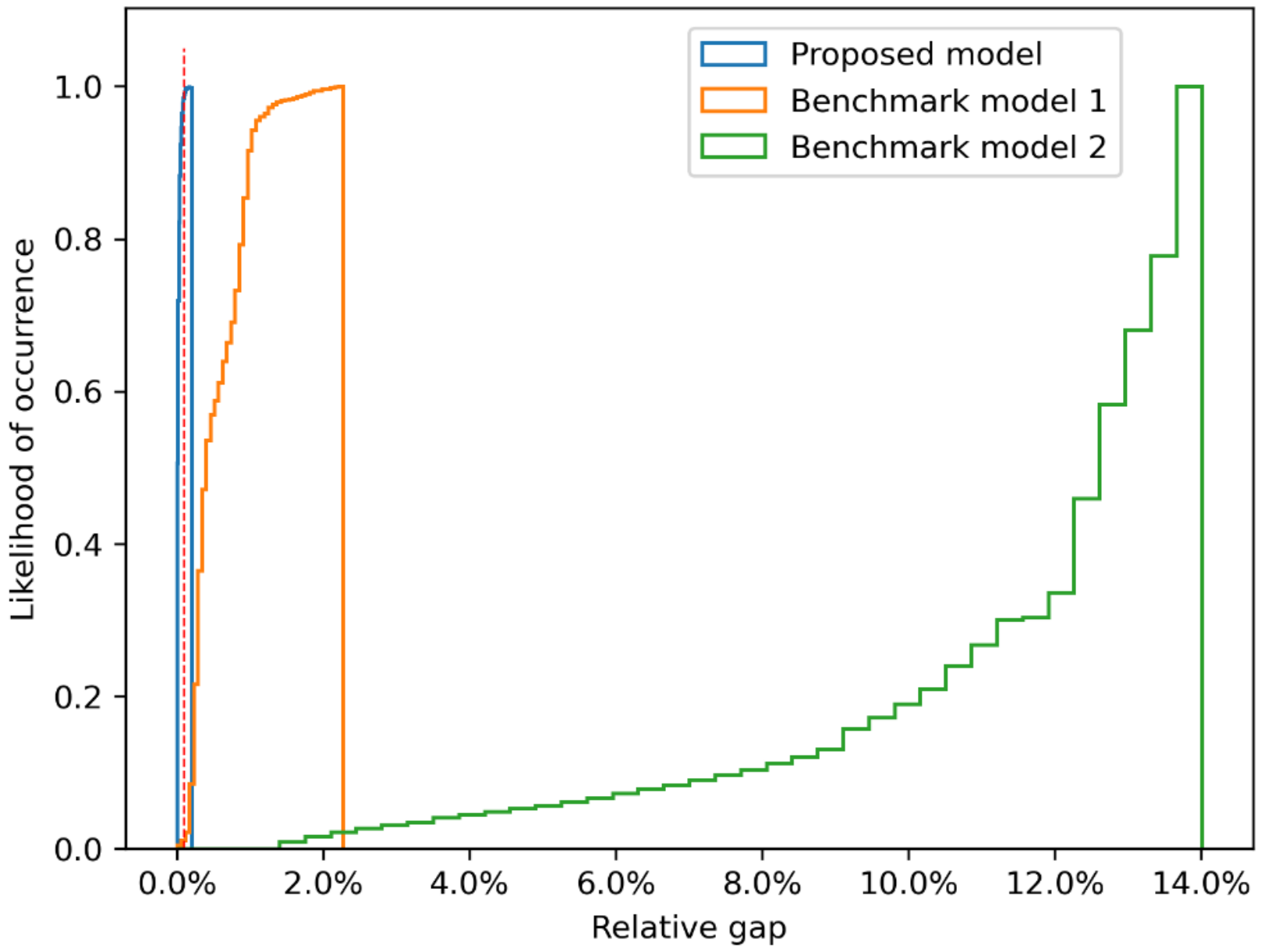

Figure 11. Cumulative histogram of corrected relative gap compared with benchmark models.

Table 1. Comparison of Average Relative Gap

| | Path 1 | Path 2 | Path 3 | Path 4 | Path 5 | Path 6 | Path 7 | Path 8 | Path 9 | Path 10 |
|---|---|---|---|---|---|---|---|---|---|---|
| Proposed model | 0.019% | 0.019% | 0.019% | 0.018% | 0.019% | 0.019% | 0.019% | 0.018% | 0.019% | 0.019% |
| Benchmark model 1 | 0.573% | 0.448% | 0.307% | 0.405% | 0.582% | 0.323% | 0.286% | 0.410% | 0.408% | 0.526% |
| Benchmark model 2 | 11.71% | 11.71% | 11.84% | 11.78% | 11.72% | 11.86% | 11.80% | 11.79% | 11.73% | 11.75% |

After analyzing the benefits of the proposed model, we proceed to investigate the impact of different ATMA routes on the traffic system travel time. The initial TSTT without ATMA vehicles is recorded as 5,656,962 hours. However, with the introduction of ATMA vehicles for maintenance, the TSTT experiences a noticeable increase. Figure 12 illustrates the increased TSTT compared to the TSTT in the absence of ATMA maintenance and the travel time of different paths. It is observed that as the path travel time increases, the additional TSTT does not necessarily increase proportionally. For instance, Path 10 results in TSTT having the largest increase of 1.25%. On the other hand, paths 2, 7, and 9 have relatively lower impacts on the total traffic system, for example, Path 9 yields the smallest increase in TSTT of 0.91%. However, the shortest path, Path 1, exerts a higher impact on the total traffic system, resulting in an increase of 1.11% in TSTT.

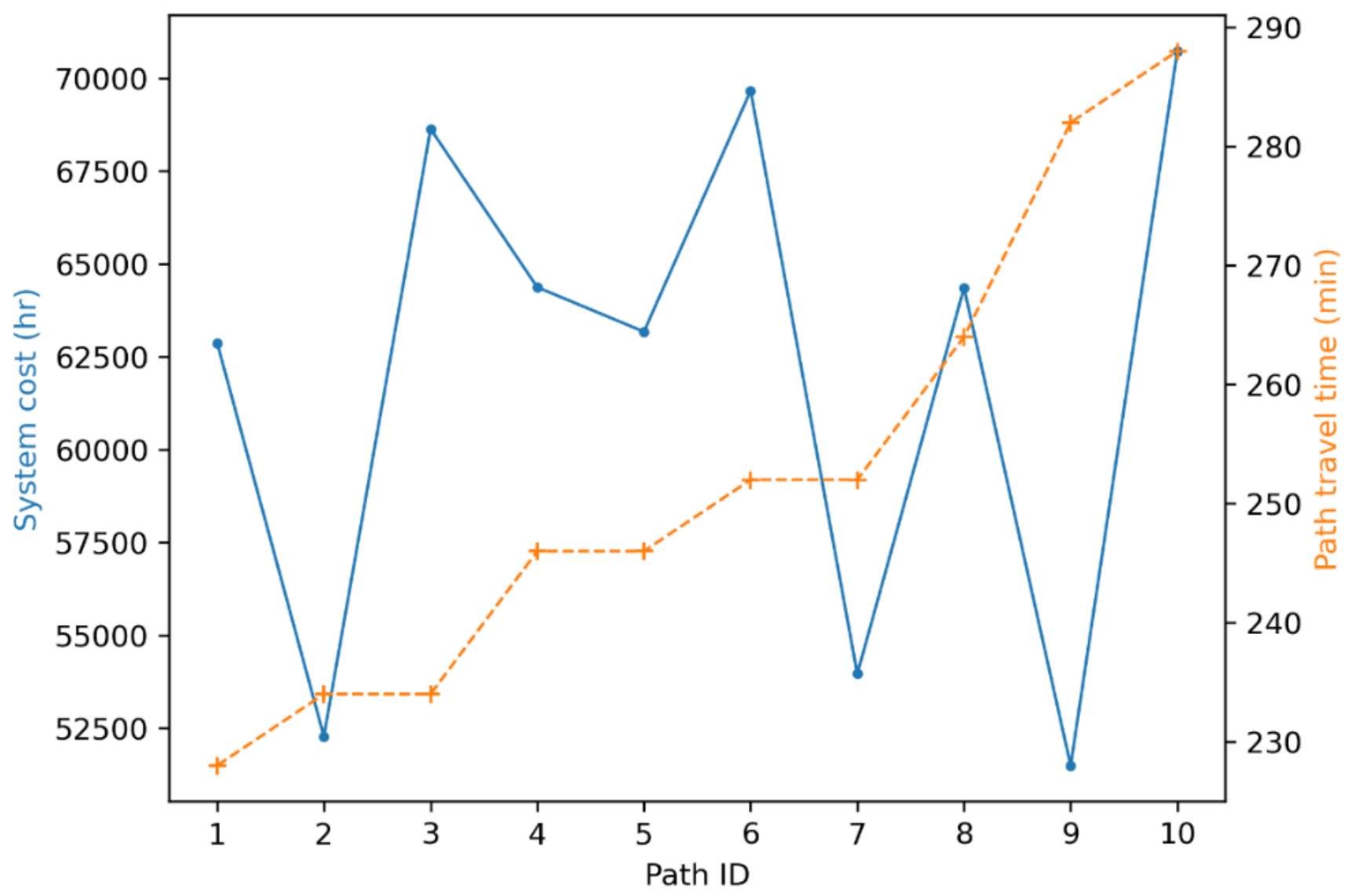

Figure 12. System cost and travel time of ten paths.

### 4.3. Sensitivity Analysis

In this section, sensitivity analysis of the Sioux Falls network is conducted to explore how the impact on the total traffic system changes when traffic demand and the speed of ATMA vehicles change.

#### *4.3.1. When Demand Changes*

The demand for each OD pair is set to increase by 10%, 20%, and 30%, respectively, to investigate the impact on total traffic system. The TSTT experiences a significant increase without ATMA vehicles as the demand rises. For instance, when the demand increases by 10%, the TSTT increases by 38%. Furthermore, with a 30% increase in demand, the TSTT shows a substantial rise of 130%. This demonstrates the high sensitivity of the TSTT to changes in demand.

Figure 13 presents the additional TSTT (or the system cost) with changing demand, illustrating that the rate of TSTT increase diminishes as demand rises. When comparing the TSTT with no ATMA vehicles, the increase in TSTT ranges from 0.77% to 1.09% for a 10% increase in demand. In the case of a 20% increase in demand, the increase in TSTT ranges from 0.68% to 0.95%. Similarly, for a 30% increase in demand, it ranges from 0.60% to 0.86%. The system costs still exhibit significant variation among different routes of ATMA vehicles. Additionally, it is worth noting that paths 9 and 2 have minimal impact on system traffic, indicating that the optimal route for ATMA vehicles remains relatively unaffected by changes in demand.

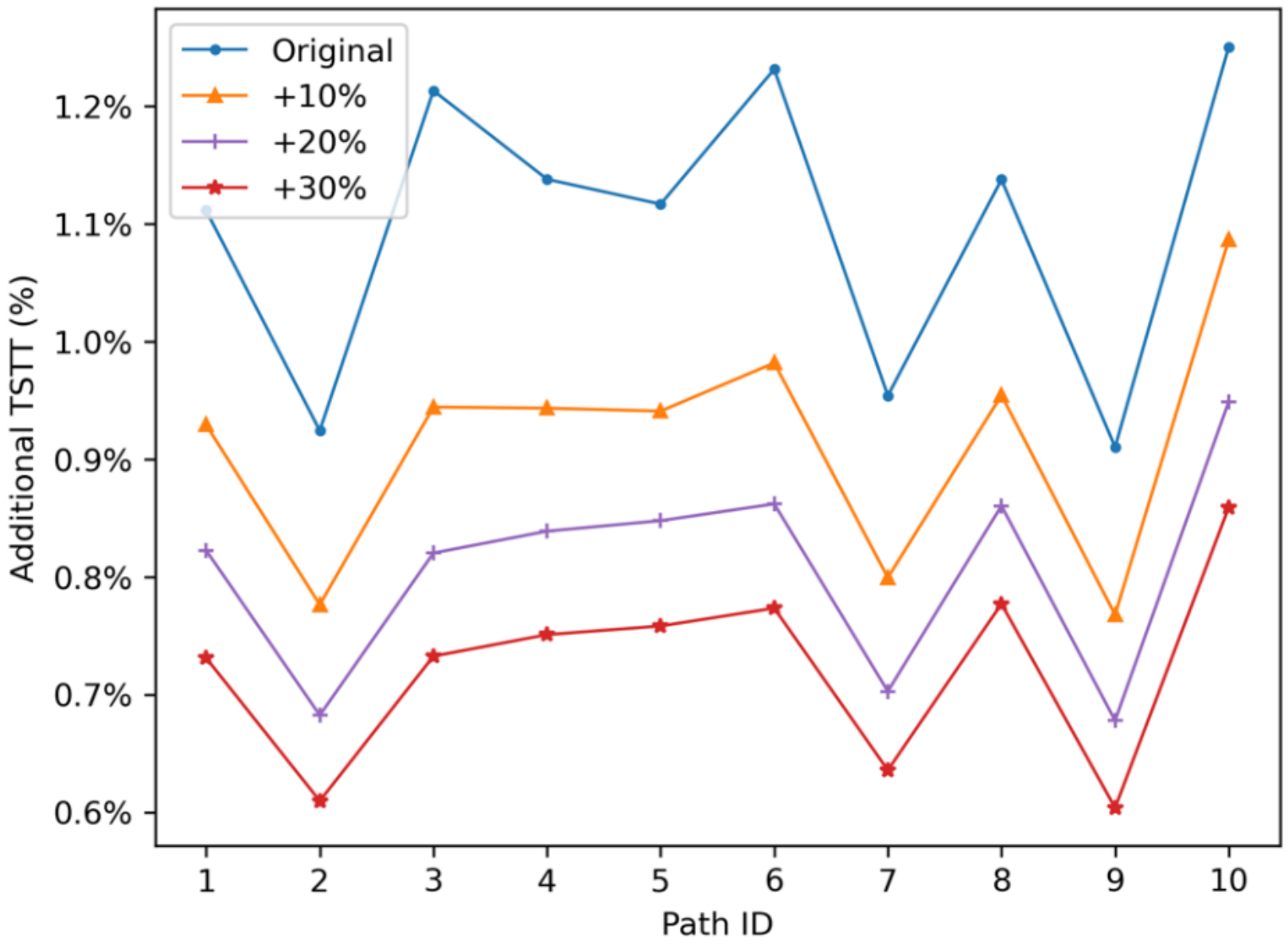

Figure 13. Additional TSTT compared with no ATMA when demand changes.

*4.3.2. When ATMA speed change changes*

We also conducted experiments to examine the impact of varying the speed of ATMA vehicles on the traffic system. Specifically, we considered speeds of 10mph, 15mph, and 20mph, and analyzed the additional TSTT for ten paths in comparison to scenarios without ATMA vehicles or capacity reduction. Figure 14 illustrates the additional TSTT with changing travel speeds. When the speed of ATMA vehicles is set at 10mph, we observe an increase in TSTT ranging from 0.91% to 1.25%. However, as the speed increases to 15mph and 20mph, the increase in TSTT ranges from 0.49% to 0.68% and 0.29% to 0.41%, respectively. These findings suggest that the rate of TSTT increase diminishes as the speed of ATMA vehicles increases. As the speed of ATMA vehicles increases, the disparity in system costs diminishes across various routes. We notice that paths 9 and 2 still have minimal impact on the system traffic, which indicates that the optimal route for ATMA vehicles remains relatively unaffected by variations in travel speed.

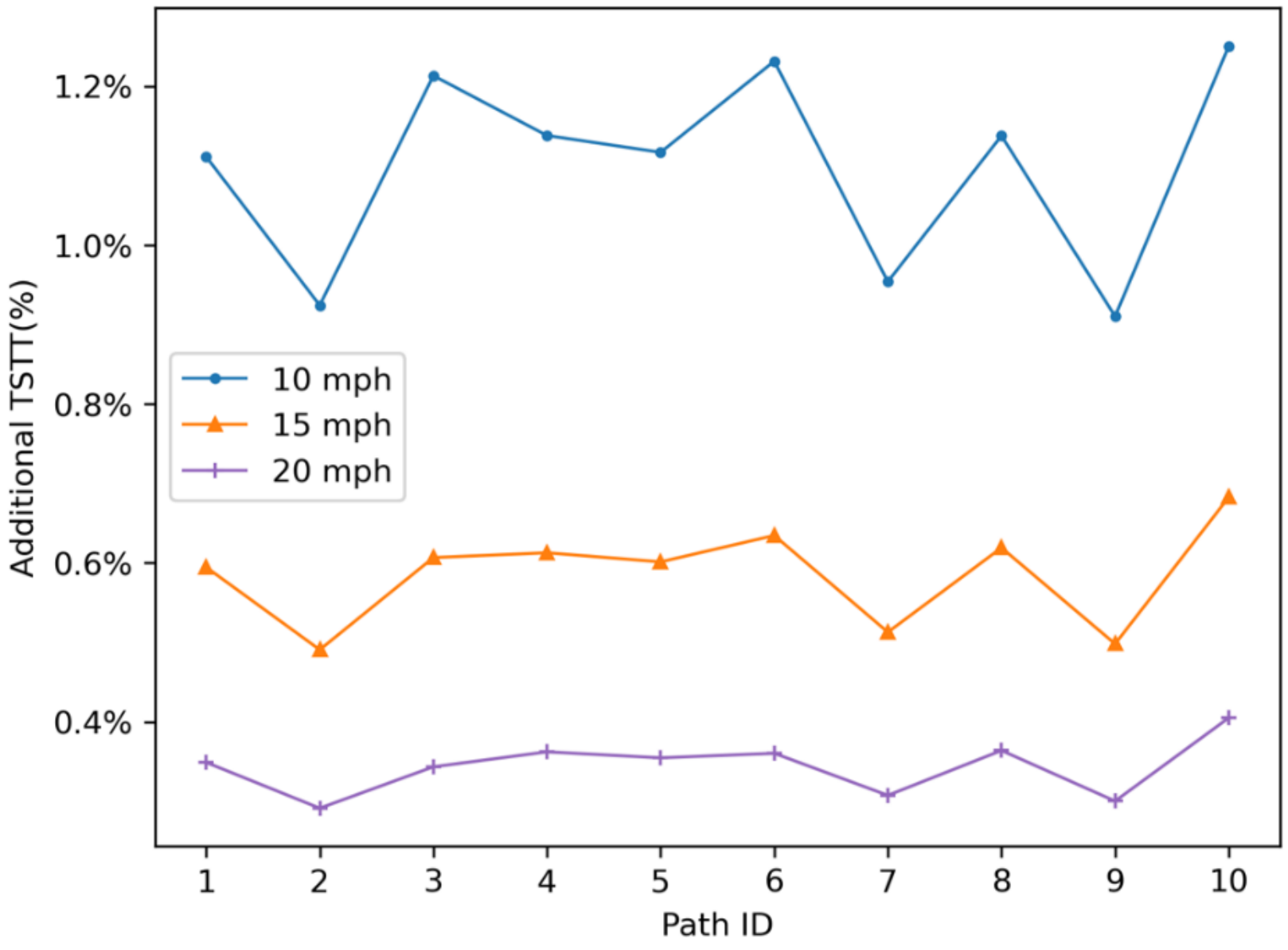

Figure 14. Additional TSTT compared with no ATMA when demand changes.

## 5. CONCLUSION

In this study, we propose a queuing-based traffic assignment approach to optimize the deployment of the ATMA system in a network. Our approach incorporates a QBTD travel time function that accounts for capacity drop, introducing dynamic characteristics into the static UETAP. We formulate the TAP by integrating the QBTD travel time function and modify a path-based algorithm to solve it, achieving user equilibrium. To validate the methodology, we conduct experiments on both a small-scale network and a larger network to investigate the convergence pattern. We compare the results with two benchmark models to analyze the benefits of time-varying capacity drop modeling and the QBTD travel time function. The corrected relative gap of the first benchmark model, which ignores capacity reduction, exhibits an average increase from 0.02% to 0.43% in the large-scale network. This highlights the significant deviation from UE when capacity drop is disregarded in the UE traffic assignment. Furthermore, when using the static BPR travel time function, there is a more notable increase in the corrected relative gap, with an average relative gap of 11.78%. Additionally, we apply our proposed queuing-based traffic assignment model to a case study focused on optimizing routing for ATMA vehicles engaged in maintenance work in work zones. We investigate and quantify the impact of different paths on the traffic system. Our findings reveal that the shortest path does not guarantee the optimal route due to the impact of the time varying capacity reduction. The result of sensitivity analysis demonstrates that system costs still exhibit significant variation among different routes of ATMA vehicles when the traffic demand increases. However, as the speed of ATMA vehicles increases, the disparity in system costs diminishes across various routes.

## 6. AUTHOR CONTRIBUTIONS

The authors confirm contribution to the paper, as follows: study conception and design: Q. Tang, X. Hu; methodology development: Q. Tang, X. Hu; analysis and interpretation of results: Q. Tang; draft manuscript preparation: Q. Tang, X. Hu. Manuscript review and editing: X. Hu. All authors reviewed the results and approved the final version of the manuscript.

## 7. APPENDICES

Appendix A – Notation List

| | |
|---|---|
| $N$ | a set of nodes |
| $A$ | a set of directed links$(i,j)$, $i \in N$ and $j \in N$ |
| $O$ | a set of origin nodes |
| $D$ | a set of destination nodes |
| $r$ | an origin node, $r \in O$ |
| $s$ | a destination node, $s \in D$ |
| $d^{rs}$ | travel demand between origin $r$ and destination $s$ |
| $\Pi^{rs}$ | a set of paths between origin $r$ and destination $s$ |
| $\pi$ | a path, $\pi \in \Pi^{rs}$ |
| $h^{\pi}$ | flow on path $\pi$ |
| $\Delta t$ | time step |
| $x_{ij}(t)$ | flow on link $(i,j)$ |
| $\delta_{ij}^{\pi}$ | the number of times link $(i,j)$ is used by path $\pi$, $\delta_{ij}^{\pi} = 1$ if path $\pi$ uses link $(i,j)$, and $\delta_{ij}^{\pi} = 0$ if it does not. |
| $tt_{ij}(Q_{ij}(t))$ | time-dependent link travel time function |
| $t_{ij}^{f}$ | free flow travel time on link $(i,j)$ |
| $Q_{ij}(t)$ | queue length on link $(i,j)$ at time $t$ |
| $w_{ij}(t)$ | delay time on link $(i,j)$ at time $t$ |
| $r_{\text{ATMA}}$ | a potential route traversing all links need to be maintained by ATMA vehicles |
| $C_{r_{\text{ATMA}}}$ | the system cost of the route $r_{\text{ATMA}}$ |
| $TSTT_0$ | the total system travel time without ATMA vehicles, following the UE principle |
| $TSTT_{r_{\text{ATMA}}}$ | the total system travel time, following the UE principle and the selection of route $r_{\text{ATMA}}$ by ATMA vehicles |

Appendix B – Derivations

Derivation of the Eq. (23):

$$\frac{\partial t_{ij}(x_{ij},t)}{\partial x_{ij}} = \frac{\partial\left(t_{ij}^{f}+\frac{Q_{ij}(t)}{\mu_{ij}(t)}\right)}{\partial x_{ij}} = \frac{\partial\left(\frac{Q_{ij}(t)}{\mu_{ij}(t)}\right)}{\partial x_{ij}} = \frac{\partial\left(\frac{\int_{t_0}^{t}\left(x_{ij}(\tau)-\mu_{ij}(\tau)\right)d\tau}{\mu_{ij}(t)}\right)}{\partial x_{ij}} = \frac{\partial(\int_{t_0}^{t}\left(x_{ij}(\tau)-\mu_0\cdot(1-\theta_{ij}(\tau)\right)d\tau)/\partial x_{ij}}{\mu_{ij}(t)}$$

$$= \frac{1}{\mu_{ij}(t)}\cdot\frac{\partial(\int_{t_0}^{t}x_{ij}(\tau)d\tau)}{\partial x_{ij}} = \frac{1}{\mu_{ij}(t)}\cdot\frac{\partial(\int_{t_0}^{t}x_{ij}(\tau)d\tau)}{\partial t}\cdot\frac{dt}{dx_{ij}} = \frac{1}{\mu_{ij}(t)}\cdot x_{ij}(t)\cdot\frac{dt}{dx_{ij}}$$

$$= \frac{x_{ij}(t)}{\mu_{ij}(t)}/\frac{dx_{ij}}{dt}$$

Derivation of the Eq. (27):

$$\frac{dt_{ij}(x_{ij},m\cdot\Delta t)}{dx_{ij}} = \frac{d\left(FFTT+w_{ij}(m\cdot\Delta t)\right)}{dx_{ij}} = \frac{d\left(FFTT+\frac{Q_{ij}(m\cdot\Delta t)}{\mu_{ij}(m\cdot\Delta t)}\right)}{dx_{ij}} = \frac{d\left(\frac{Q_{ij}(m\cdot\Delta t)}{\mu_{ij}(m\cdot\Delta t)}\right)}{dx_{ij}} = \frac{1}{\mu_{ij}(m\cdot\Delta t)}\cdot\frac{dQ_{ij}(m\cdot\Delta t)}{dx_{ij}}$$

$$= \frac{1}{\mu_{ij}(m\cdot\Delta t)}\cdot\frac{d(\int_{(m-1)\cdot\Delta t}^{m\cdot\Delta t}\left(x_{ij}-\mu_{ij}(\tau)\right)d\tau)}{dx_{ij}} = \frac{1}{\mu_{ij}(m\cdot\Delta t)}\cdot\frac{d\int_{(m-1)\cdot\Delta t}^{m\cdot\Delta t}x_{ij}d\tau}{dx_{ij}}$$

$$= \frac{1}{\mu_{ij}(k\cdot\Delta t)}\cdot\frac{d\left(x_{ij}\cdot(k\cdot\Delta t-t_0-(k-1)\cdot\Delta t)\right)}{dx_{ij}} = \frac{\Delta t}{\mu_{ij}(k\,\Delta t)}$$